\documentstyle[editedvolume]{crckapb} 

\input psfig.tex

\newcommand{\be}{\begin{equation}}
\newcommand{\ee}{\end{equation}}
\def\lsim{\lower.5ex\hbox{$\; \buildrel < \over \sim \;$}}
\def\gsim{\lower.5ex\hbox{$\; \buildrel > \over \sim \;$}}

\begin{opening}
\title{Computation of Mass-Outflow Rates From Advective Accretion Disks Around Black Holes }

\author{Tapas K. Das }
\institute{S. N. Bose National Centre For Basic Sciences\protect\\
JD Block, Salt Lake, Sector-III, Calcutta-700091,\protect\\ 
India\protect\\
e-mail:tdas@boson.bose.res.in
}

\end{opening}

\runningtitle{Outflows from Advective Disks}

\begin{document}
\noindent Appearing in `Observational Evidence for Black Holes in the Universe', Ed.
S.K. Chakrabarti (Kluwer Academic Publishers: Holland), p. 113, (1998).


\section{Introduction}

The existing models which study the origin, acceleration and collimation of mass outflow
in the form of jets from AGNs and Quasars are roughly of three types. The first type of
solutions confine themselves to the jet properties only, completely
decoupled from the internal properties of accretion disks (see, e.g., Begelman, Blandford \& Rees, 1984).
In the second type, efforts are made to correlate the internal disk structure with that of the
outflow using both hydrodynamic (e.g., Chakrabarti 1986)
and magnetohydrodynamic considerations (K\"onigl 1989; Chakrabarti \&
Bhaskaran 1992). In the third type, numerical simulations are carried out
to actually see how matter is deflected from the equatorial plane 
towards the axis (e.g., Hawley, Smarr \& Wilson, 1984, 1985; 
Eggum, Katz \& Coroniti, 1985; Molteni, Lanzafame \& Chakrabarti, 1994, hereafter MLC94; 
Molteni, Ryu \& Chakrabarti, 1996, hereafter MRC96; Ryu, Chakrabarti \& Molteni, 1997; Nobuta \& Hanawa, 1998). 
>From the analytical front, although the wind type solutions and accretion type solutions
come out of the same set of governing equations (Chakrabarti 1990), there are few, 
and mostly qualitative attempt to find connections among them (Chakrabarti, 1997a). 
As a result, the estimation of the outflow rate from the inflow rate has been difficult.
Our work, {\it for the first time}, quantitatively connects the topologies of the inflow and the
outflow. The simplicity of black holes and neutron stars lie in the fact 
that they do not have atmospheres. But the disks surrounding them have, and 
similar method as employed in stellar atmospheres should be applicable to the disks.
Our approach in this paper is precisely this. We first
determine the properties of the rotating inflow and outflow
and identify solutions to connect them. In this manner, we
self-consistently determine the mass outflow rates. 

Before we present our results, we describe basic properties of
the rotating inflow and outflow. As is well known (MLC94, MRC96 and references therein), 
in the centrifugal pressure supported boundary layer (CENBOL) 
the flow becomes hotter and denser and for all practical purposes 
behaves as the stellar atmosphere so far as the formation of 
outflows are concerned. In case where the shock does not form, 
regions around pressure maximum achieved just outside the inner sonic 
point would also drive the flow outwards.
We calculate the mass outflow rates (${{R}_{\dot m}}$)  rate as a function of the inflow parameters, such
as specific  energy and angular momentum, accretion rate, polytropic index etc.
We explore both the polytropic and the isothermal outflows. 
A detailed report on this type of outflows is presented elsewhere (Das \& Chakrabarti,
1998, hereafter DC98).

The plan of this paper is the following: In the next Section, we describe our model
and present the governing equations for the inflow and outflow. We also provide the 
solution procedure of those equations. In \S 3, we  present results of our computations. 
Finally, in \S 4, we draw our conclusions.

\section{Model Description, Governing Equations and the solution Procedure}

We consider thin, axisymmetric polytropic inflows in vertical equilibrium 
(otherwise known as 1.5 dimensional flow). We 
ignore the self-gravity of the flow and viscosity is assumed to be significant only
at the shock so that entropy is generated. We do the calculations using Paczy\'nski-Wiita (1980)
potential which mimics surroundings of the Schwarzschild black hole. 
The equations (in dimensionless units) governing the inflow are:\\
$$
{\cal E}=\frac{{u_e}^2}{2} +n{{a_e}^2}+
\frac{{\lambda}^2}{2{r^2}}-\frac{1}{2(r-1)} .
\eqno{(1)}
$$
$$
{{\dot M}_{in}}={u_e}{{\rho}_e}r{h_e}(r),
\eqno{(2)}
$$
(For detail, see, Chakrabarti, 1989 hereafter C89).
The equations governing the polytropic outflow are
$$
{\cal E}=\frac{\vartheta^2}{2}+{n^\prime}{{a_e}^2}+\frac{{\lambda}^2}{2{{r_m}^2(
r)}}
-\frac{1}{2(r-1)}
\eqno{(3)}
$$
And
$$
{{\dot M}_{out}}={\rho}{\vartheta}{\cal A}(r).
\eqno{(4)}
$$
Where ${r_m}$ is the mean axial distance of the flow and ${{\cal A}(r)}$
is the cross sectional area through which mass is flowing out. (For detail,
see, DC98.) $\gamma$ of the outflow was taken to be smaller than that of the inflow
because of momentum deposition effects.
The outflow angular momentum $\lambda$ is chosen to be the same as in the
inflow, i.e., no viscous dissipation is assumed to be present in the inner region of the
flow close to a black hole. Considering that viscous time scales are longer compared to 
the inflow time scale, it may be a good  assumption in the disk, but it may not be
a very good assumption for the outflows which are slow prior to the acceleration and
are therefore, prone to viscous transport of angular momentum.  
Detailed study of the outflow rates in presence of viscosity 
and magnetic field is in progress and would be presented elsewhere.
The Isothermal outflow is governed by the following equations
$$
\frac{{\vartheta_{iso}}^2}{2}+{{C_s}^2}ln{\rho}+\frac{\lambda^2}{2{{r_m}(r)}^2}
-\frac{1}{2(r-1)}={\rm Constant}
\eqno{(5)}
$$
And
$$
{{\dot{M}}_{out}}=\rho {\vartheta_{iso}}{{\cal A}(r)}.
\eqno{(6)}
$$
Here, the area function remains the same above. A subscript {\it  iso} of 
velocity ${\vartheta}$ is kept to distinguish from the velocity in 
the polytropic case. This is to indicate the 
velocities are measured here using completely different assumptions. For details, see DC98.

In both the models of the outflow, we assume that the flow is primarily radial. Thus the
$\theta$-component of the velocity is ignored ($\vartheta_\theta  << \vartheta$).

\subsection{Procedure to solve for disks and outflows simultaneously}

For polytropic outflows, we solve equations (1-4) simultaneously using numerical
techniques (for detail, see, DC98).
In this case the specific energy ${\cal E}$ is assumed to  
remain fixed throughout the flow trajectory as it moves from the disk to the jet. 
At the shock, entropy is generated
and hence the outflow is of higher entropy for the same specific energy.

A supply of parameters ${\cal E}$, $\lambda$, $\gamma$ and $\gamma_{o}$ makes
a self-consistent computation of $R_{\dot m}$ possible when the shock is present. 
In the case where the shocks do not form, the procedure is a bit different. 
It is assumed that the maximum amount of matter comes out from the 
place of the disk where the thermal pressure of the inflow attains its maximum 
and the outflow is assumed to have the same quasi-conical shape with annular
cross-section ${\cal A} (r)$ between the funnel wall and the centrifugal barrier as already defined.
For this case, the compression ratio of the gas at the pressure maximum between the inflow and outflow
$R_{comp}$ is supplied as a free parameter, since it may be otherwise very difficult to 
compute satisfactorily. In the presence of shocks, such problems do not arise as the
compression ratio is obtained self-consistently. For isothermal outflow, 
it is assumed that the outflow has exactly the {\it same} temperature as that of the
post-shock flow, but the energy is not conserved as matter goes from disk to the 
wind. The polytropic index of the inflow can vary but that of the outflow is
always unity. The other assumptions and logical steps are exactly same
as those of the case where the outflow is polytropic. Here we solve 
equations (1-2) and (5-6) simultaneously using numerical technique to get 
results. (For details, see, DC98). 

\section{Results}

\subsection{Polytropic outflow coming from the post-shock accretion disk}

\begin{figure}
\vbox{
\vskip 0.0cm
\hskip 11.5cm
\centerline{
\psfig{figure=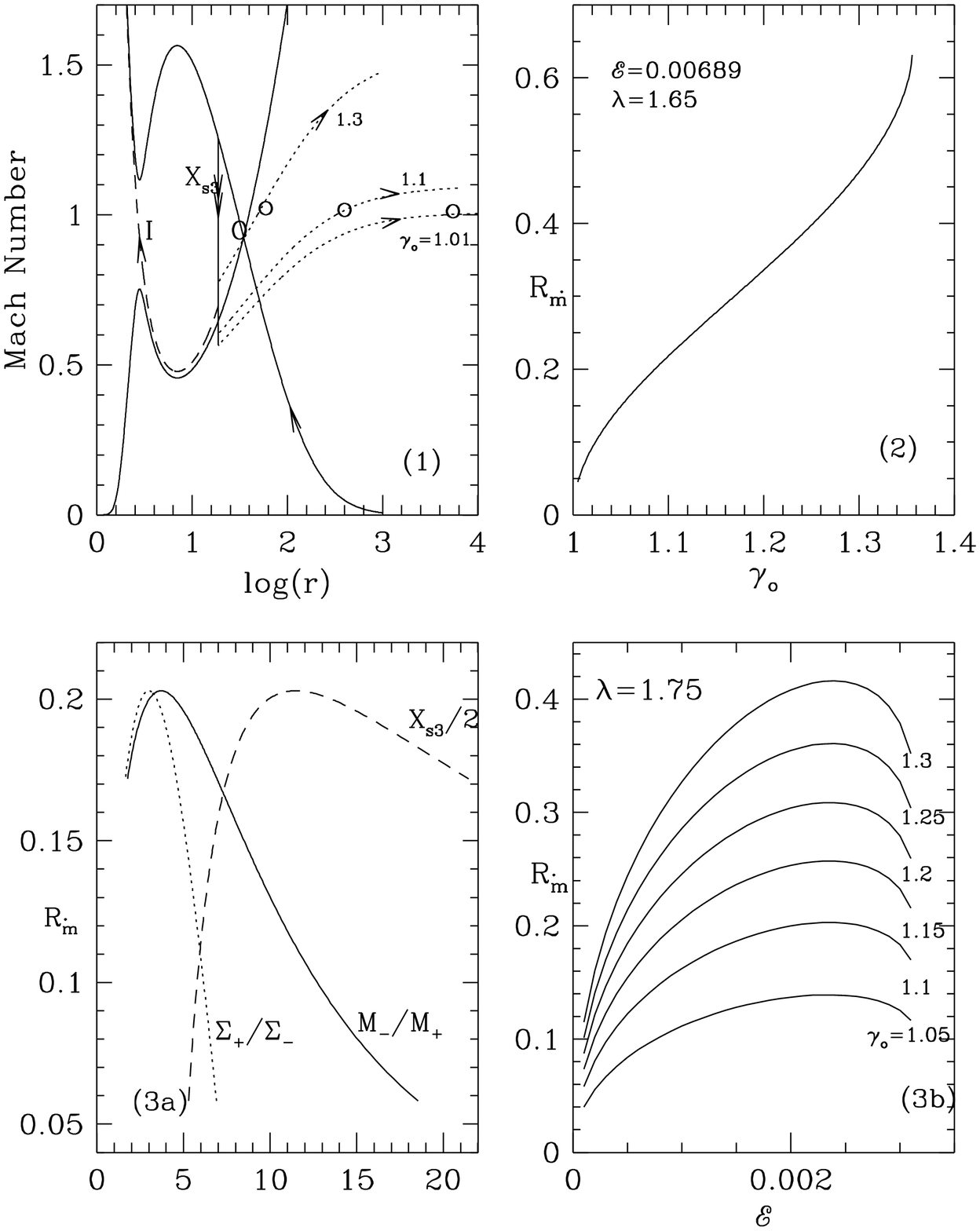,height=13truecm,width=13truecm,angle=0}}}
\vskip 0.5cm
\noindent{\small {\bf Fig. 1-3}: Mach number of the flow is plotted against logarithmic radial
distance both for the inflow and outflow (Fig. 1). The ratio of mass outflow rate and mass inflow rate
is plotted against the polytropic index of the outflow (Fig. 2). The same ratio is plotted against
the shock strength, shock location and the ratio of the integrated density  (Fig. 3a) and specific
energy ${\cal E}$ and polytropic index of the outgoing flow $\gamma_o$ (Fig. 3b). See text for details.}

\end{figure}

Figure 1 shows a typical solution which combines the accretion and the outflow. The input parameters
are ${\cal E}=0.00689$, ${\lambda=1.65}$ and $\gamma=4/3$ corresponding to 
relativistic inflow (see Fig. 5 and 6 of C89). The solid curve with an arrow represents 
the pre-shock region of the inflow and the long-dashed curve represents the post-shock
inflow which enters the black hole after passing through the inner sonic point (I).
The solid vertical line at $X_{s3}$ with double arrow represents the shock transition.
Three dotted curves represent three outflow solutions for the parameters 
$\gamma_{o}=1.3$ (top), $1.1$ (middle) and $1.03$ (bottom). The outflow
branches shown pass through the corresponding sonic points. It is
evident from the figure that the outflow  moves along solution curves which are completely different from that
of the `wind solution' of the inflow which passes through the outer sonic point `O'.
The mass loss ratio $R_{\dot m}$ in these cases are $0.47$, $0.22$ and $0.06$ 
respectively. Figure 2 shows the ratio $R_{\dot m}$ as $\gamma_{o}$ is varied. Only the 
range of $\gamma_{o}$ for which the shock-solution is present is shown here. 
In Fig. 3a we show the variation of the ratio $R_{\dot m}$ 
of the mass outflow rate inflow rate as a function of 
the shock-strength (solid) $M_-/M_+$ (Here, $M_-$ and $M_+$ 
are the Mach numbers of the pre- and post-shock flows respectively.),
the compression ratio (dotted) $\Sigma_+/\Sigma_-$ (Here, $\Sigma_-$ and $\Sigma_+$ 
are the vertically integrated matter densities in the pre- and post- shock
flows respectively), and the stable shock location (dashed) $X_{s3}$ (in the notation of C89).
Other parameters are $\lambda=1.75$ and $\gamma_{o} = 1.1$. 
Note that the ratio $R_{\dot m}$ does not peak near the strongest 
shocks! Shocks are stronger when they are located closer to the black 
hole, i.e., for smaller energies. 
In Fig. 3b where $R_{\dot m}$ is plotted 
as a function of the specific energy ${\cal E}$ (along x-axis) 
and $\gamma_{o}$ (marked on each curve). Specific angular momentum 
is chosen to be $\lambda=1.75$ as before. The peak in $R_{\dot m}$ 
is  observed (see also, Chakrabarti, 1997a, and Chakrabarti, this volume).
To have a better insight of the behavior of the outflow we plot in Fig. 4 $R_{\dot m}$
as a function of the polytropic index of the incoming flow $\gamma$. 
The range of $\gamma$ shown is the range for which shock forms in the flow. We also plot the
variation of velocity $\vartheta_o$, density $\rho_o$ and area ${\cal A}(r)$ of the outflow
at the location where the outflow leaves the disk.
These quantities are scaled from the corresponding dimensionless units as $\vartheta_o \rightarrow 
2 \times 10^4 \vartheta_o -558$, $\rho_o \rightarrow 10^{22} \rho_o$ and ${\cal A} 
\rightarrow 0.0005 {\cal A} $ respectively in order to bring them in the
same scale. The non-monotonic nature of the variation of $R_{\dot m}$ with $\gamma$ is observed.

\subsection{Polytropic outflow coming from the region of the maximum pressure}

\begin{figure}
\vbox{
\vskip 0.0cm
\hskip 11.5cm
\centerline{
\psfig{figure=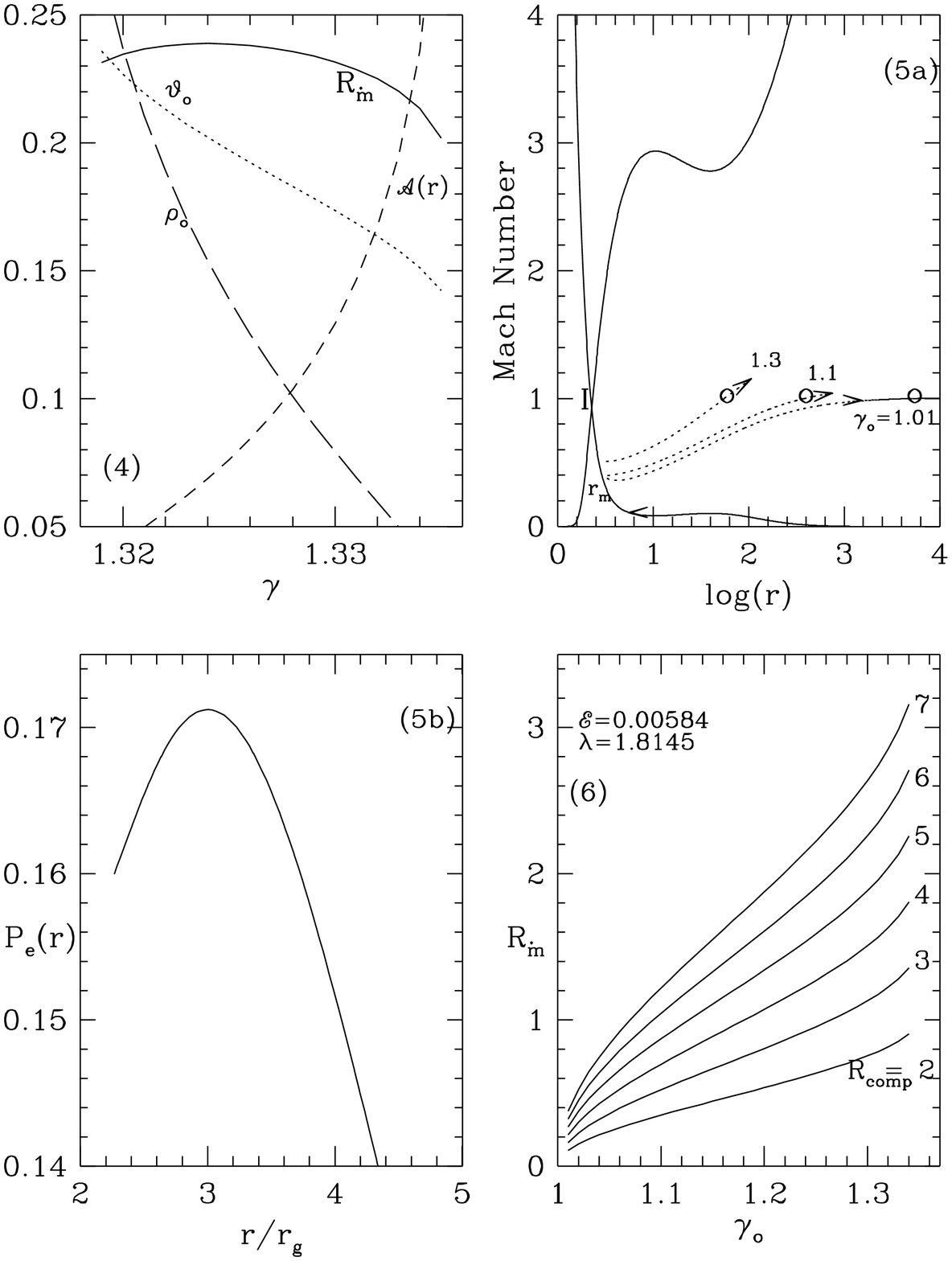,height=13truecm,width=13truecm,angle=0}}}
\vskip 0.5cm
\noindent{\small {\bf Fig. 4-6}: Variation of velocity, density, cross sectional 
area and the rate ratio as a function of the polytropic index of the inflow (Fig. 4).
Variation of the Mach number for inflow and outflow when shocks are not present (Fig. 5a).
Thermal pressure variation as a function of the radial distance $r/r_g$ showing
a distinct maximum (Fig. 5b). Variation of $R_{\dot m}$ when both the
compression ratio at the pressure maxima and polytropic index of the outflow
are changed (Fig. 6). See text for details.}
\end{figure}

In this case, the inflow parameters are chosen from region {\bf I} (see C89)
so that the shocks do not form. Here, the inflow passes through 
the inner sonic point only. The outflow is assumed to be coming 
out from the regions where the polytropic inflow has maximum pressure. 
Figure 5a shows a typical solution. The arrowed solid curve shows the inflow 
and the dotted arrowed curves show the outflows 
for $\gamma_o=1.3$ (top), $1.1$ (middle) and $1.01$ (bottom). 
The ratio $R_{\dot m}$ in these cases 
is given by $0.66$, $0.30$ and $0.09$ respectively. 
The specific energy and angular momentum are chosen to be ${\cal E}=0.00584$
and $\lambda=1.8145$ respectively.
The pressure maximum occurs outside the inner sonic point at $r_m$ when 
the flow is still subsonic. Figure 5b shows the variation of 
thermal pressure of the flow with radial distance. The peak is clearly visible. 
Figure 6 shows the 
ratio $R_{\dot m}$ as a function of $\gamma_{o}$ for various choices of the
compression ratio $R_{comp}$ of the outflowing gas at the pressure maximum: $R_{comp}=2$ for the
bottom curve and $7$ for the top curve. Note that flows with highest compression ratios
produce highest outflow rates, evacuating the disk which is responsible
for the quiescent states in X-ray Novae systems and also in some systems with massive black holes
(e.g., our own galactic centre?). 

The location of maximum pressure being close to the black hole, it  may be 
very difficult to generate the outflow from this region. Thus, it is expected that the
ratio $R_{\dot m}$ would be larger when the maximum pressure is located farther out.
This is exactly what we see in Fig. 7, where we plot $R_{\dot m}$ against the location of
the pressure maximum (solid curve). Secondly, if our guess that the outflow rate could be related to the
pressure is correct, then the rate should increase as the pressure at the maximum rises.
That's also what we observe in Fig. 7. We plot $R_{\dot m}$ as a function of the 
actual pressure at the pressure maximum (dotted curve). The mass loss is found to be a strongly
correlated with the thermal pressure. Here we have multiplied non-dimensional thermal 
pressure by $1.5 \times 10^{24}$ in order to bring them in the same scale.

\subsection{Isothermal outflow coming from the post-shock accretion disk}

\begin{figure}
\vbox{
\vskip 0.0cm
\hskip 11.5cm
\centerline{
\psfig{figure=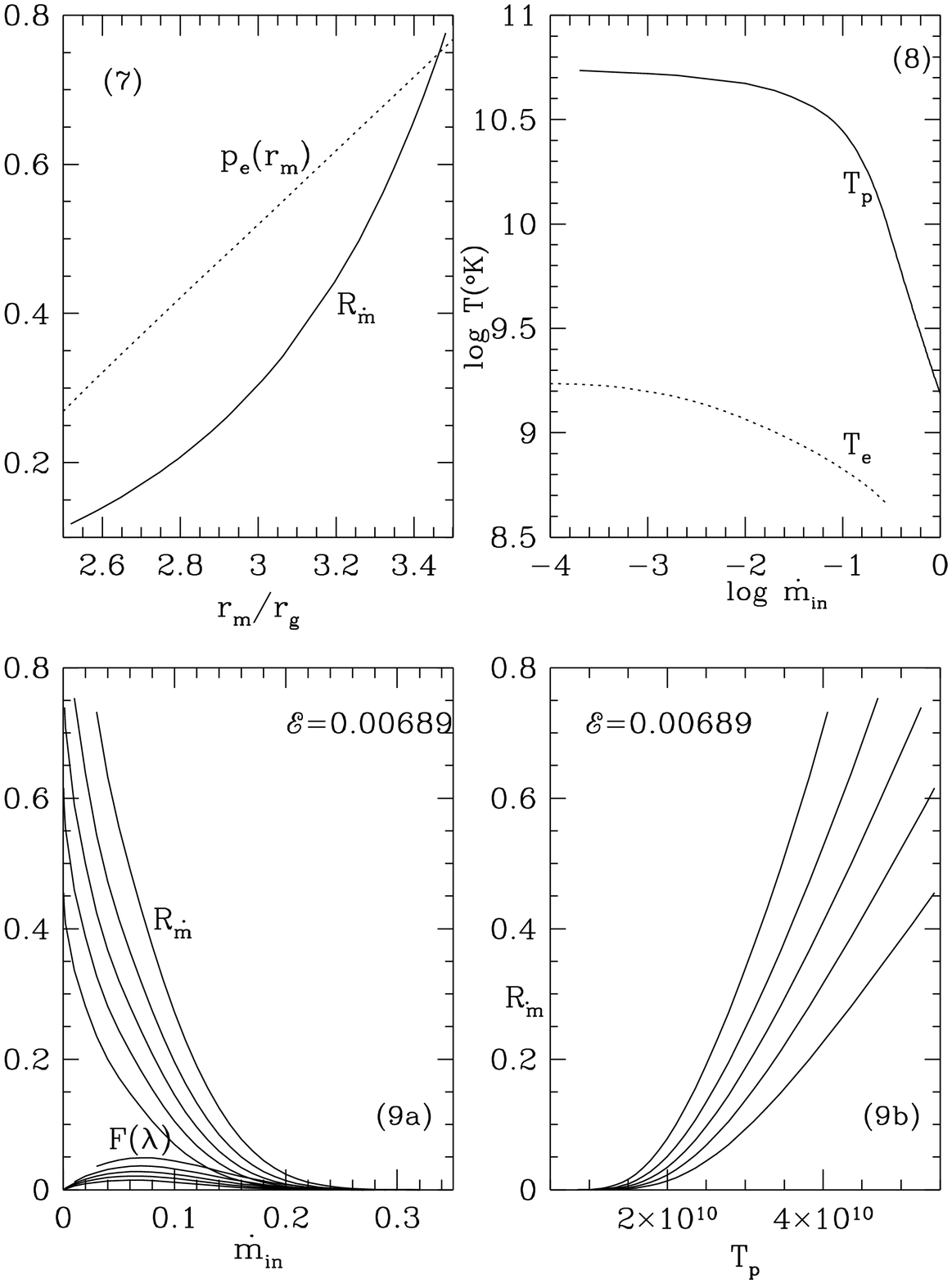,height=13truecm,width=13truecm,angle=0}}}
\vskip 0.5cm
\noindent{\small {\bf Fig. 7-9}: Variation of the maximum pressure and $R_{\dot m}$
with the location where the pressure maxima occur (Fig. 7). Proton and electron
temperatures in the advective region as a function of the inflow disk accretion rate
${\dot m}_{in}$ (Fig. 8). Variation of the $R_{\dot m}$ and angular momentum
flux $F(\lambda)$ as a function of the accretion rate of the inflow (Fig. 9a). Variation of
$R_{\dot m}$ with proton temperature $T_p$ (Fig. 9b). See text for details.}

\end{figure}

Here the temperature of the outflow
is obtained from the proton temperature of the advective region of the disk. The proton temperature
is obtained using the Comptonization, bremsstrahlung, inverse bremsstrahlung and Coulomb processes
(Chakrabarti, 1997b and references therein). Figure 8 shows the effective
proton temperature and the electron temperature of the post-shock advective region as a function of the
accretion rate (in logarithmic scale)  of the Keplerian component of the disk. 
In Fig. 9a, we show the ratio $R_{\dot m}$ as a function of the Eddington rate of the
incoming flow for a range of the specific angular momentum. In the low luminosity
objects the ratio is larger. Angular momentum is varied from $\lambda=1.63$ (top curve) 
to $1.65$ (bottom curve). An interval of $\lambda=0.005$ was used.
The ratio is very sensitive to the angular momentum since it changes the
shock location rapidly and therefore changes the post-shock temperature very much. 
We also plot the outflux of angular momentum $F ({\lambda})=\lambda {\dot m}_{in} R_{\dot m}$
which has a maximum at intermediate accretion rates. In dimensional units, these
quantities represent significant fractions of angular momentum  of the entire disk
and therefore the rotating outflow can help accretion processes. Curves are drawn for different $\lambda$
as above. In Fig. 9b, we plot the  variation of the ratio directly with the proton temperature of the
advecting region. The outflow is clearly thermally driven. Hotter flow 
produces more winds as is expected. The angular momentum associated
with each curve is same as before.

\subsection{Isothermal outflow coming from the region of the maximum pressure}

This case produces very similar result as in the above case, except that like Section 3.2 the
outflow rate becomes more than a hundred percent of the inflow rate when the proton temperature is very high. 
This phenomenon may be responsible for producing quiescent states in some black hole
candidates.

\section{Conclusions}

The basic conclusions of this paper are the followings:\\

\noindent a) It is possible that most of the outflows are coming from the centrifugally 
supported boundary layer (CENBOL) of the accretion disks.\\
\noindent b) The outflow rate generally increases with the proton temperature of CENBOL. In other
words, winds are, at least partially, thermally driven. 
This is reflected more strongly when the outflow is isothermal.\\
\noindent c) Even though specific angular momentum of the flow increases the size of the CENBOL,
and one would have expected a higher mass flux  in the wind, we find that the rate of the
outflow is actually anti-correlated with the $\lambda$ of the inflow. This is because the
average proton temperature of CENBOL goes down with $\lambda$. \\
\noindent (d) Presence of 
significant viscosity in CENBOL may reduce angular momentum of the outflow.  When this 
is taken into account, we find that the rate of the outflow is correlated with  
$\lambda$ of the outflow. This suggests that the outflow is partially centrifugally driven as well.\\
\noindent e) The ratio $R_{\dot m}$ is generally anti-correlated with the inflow accretion rate. That is,
disks of lower luminosity would produce higher ratio $R_{\dot m}$.\\
\noindent f) Generally speaking, supersonic region of the inflow do not have pressure maxima. Thus,
outflows emerge from the subsonic region of the inflow, whether the shock actually forms or not.\\

An interesting situation arises when the polytropic index of the outflow is large 
and the compression ratio of the flow is also very high. In this case, the flow virtually
bounces back as the winds and the outflow rate can be temporarily larger compared with the
inflow rate, thereby evacuating the disk. In this range of parameters, most, if not all,
of our assumptions breakdown completely because the situation becomes inherently time-dependent.
It is possible that some of the black hole systems, including that in our own galactic centre,
may have undergone such evacuation phase in the past and gone into quiescent phase. 

So far, we made the computations around a Schwarzschild black hole. 
The mass outflow rates for Kerr black holes
are being studied and the results would be reported elsewhere (Das, 1998a).
For the quasi-spherical Bondi-type accretion onto black holes, 
the accretion disk does not form, the freely falling matter may 
produce a standing collisionless shock due to the plasma instabilities
and the nonlinearity introduced in the flow due to even a small
density perturbation. Outflow from these cases is also being studied
and would be reported elsewhere (Das, 1998b).\\
We made a few assumptions, some of which may be questionable. 
Nevertheless, we believe that our calculation 
is sufficiently illustrative and gives a direction which can be followed in the future.

\end{document}